\documentclass[preprint,12pt,authoryear]{elsarticle}

\usepackage{amsmath}
\usepackage{algorithmic}
\usepackage{algorithm}
\usepackage{hyperref}
\usepackage{enumitem}
\setlist[itemize]{label=-}
\usepackage{amssymb}
\usepackage{multirow}
\usepackage{booktabs}
\usepackage{siunitx}
\usepackage{tabularx}
\usepackage{rotating}
\usepackage{geometry}
\usepackage{amsmath}
\usepackage{changepage}
\usepackage{physics}
\usepackage{subcaption} 
\usepackage{graphicx} 
\usepackage{doi}
\usepackage{float}
\usepackage{booktabs} 
\usepackage{hyperref}
\usepackage{natbib}
\journal{Nuclear Physics B}
\usepackage{grffile}  % Permette spazi e caratteri speciali nei nomi

\newtheorem{exemple}{Exemple}

\begin{document}

\thispagestyle{plain}

\begin{center}
    \LARGE{\textbf{Quantum Adaptive Search: A Hybrid Quantum-Classical Algorithm for Global Optimization of Multivariate Functions}}
\end{center}

\begin{center}
\vspace{4pt}
\large
    G. Intoccia\textsuperscript{1}, U. Chirico \textsuperscript{1,2}, V. Schiano Di Cola \textsuperscript{2}, G. Pepe\textsuperscript{1}, S. Cuomo \textsuperscript{1} \\
    
\vspace{0.4cm}
\small
   \textsuperscript{1} University of Naples Federico II \textsuperscript{2} Quantum2pi S.r.l. 

\end{center}

\begin{small}
\begin{center}
\vspace{9pt}
\textbf{Abstract}    
\end{center}

\begin{adjustwidth}{20pt}{20pt}
\small \noindent 
This work presents Quantum Adaptive Search (QAGS), a hybrid quantum-classical algorithm for the global optimization of multivariate functions. The method employs an adaptive mechanism that dynamically narrows the search space based on a quantum-estimated probability distribution of the objective function. A quantum state encodes information about solution quality through an appropriate complex amplitude mapping, enabling the identification of the most promising regions, and thus progressively tightening the search bounds; then a classical optimizer performs local refinement of the solution. The analysis demonstrates that QAGS ensures a contraction of the search space toward global optima, with controlled computational complexity. The numerical results on the benchmark functions show that, compared to the classical methods, QAGS achieves higher accuracy while offering advantages in both time and space complexity.
\end{adjustwidth}

\end{small}

\section{Introduction}

Quantum optimization algorithms aim to utilize quantum mechanical principles to solve problems that are classically hard due to their exponential scaling in the number of variables. Variational methods such as the Variational Quantum Eigensolver (VQE) and the Quantum Approximate Optimization Algorithm (QAOA) have gained popularity. QAOA, through the alternating application of cost and mixing Hamiltonians, explores the solution space in a non-classical manner, while a classical optimizer iteratively refines the parameters to maximize the expectation value of the target Hamiltonian. As demonstrated in the original work \cite{farhi2014}, this approach can offer theoretical advantages for problems like MaxCut, where quantum superposition allows sampling states that would be difficult to explore classically. Despite promising theoretical speedups, real-world implementations of quantum algorithms have so far failed to demonstrate significant advantages over classical methods, except in highly specialized scenarios and with massive resource overhead. Indeed, it has been shown that quantum advantage for Max-Cut requires hundreds of qubits \cite{Guerreschi}, and even when QAOA exhibits accelerations, these are often negated by the overhead of error correction and the need for repeated measurements.
Benchmarks \cite{Shaydulin_2019} show QAOA’s NISQ implementations achieve only marginal improvements, with performance highly sensitive to parameters and problem structure, limiting its current scalability.

In this work, we introduce a non-variational approach to the problem of optimizing a continuous n-variable function. Specifically, we propose a hybrid non-variational method that leverages amplitude encoding, serving as an alternative to the aforementioned quantum optimization algorithms. This approach may offer computational advantages over current classical algorithms, particularly for large-scale problems and the search for minima of complex functions in high-dimensional domains.
The QAGS algorithm effectively combines the advantages of quantum computing with classical optimization techniques, and its implementation enables overcoming the curse of dimensionality.

\section{Method Implementation }
We consider the problem of finding the global minimum of a function $f: \Omega \rightarrow \mathbb{R}$ defined on a $d$-dimensional domain.

\begin{equation}
\mathbf{x}^* = \arg\min_{\mathbf{x} \in \Omega} f(\mathbf{x}) \hspace{0.8cm} \Omega = \prod_{i=1}^d [l_i, u_i]
\end{equation}

where $[l_i, u_i]$ are the initial bounds for each variable $x_i$.

The algorithm start defining a gird on the initial space. For each dimension $i$, we discretize the interval $[l_i, u_i]$ in $2^n$ points, where $n$ is the number of qubits for each dimension:

\begin{equation}
x_{ij} = l_i + j\frac{u_i-l_i}{2^n-1}, \quad j=0,\ldots,2^n-1
\end{equation}

This discretization creates a uniform grid that maps classical points to quantum states.
We construct a quantum state whose amplitudes encode the value of the objective function.

\begin{equation}
\psi(\mathbf{x}) = \frac{1}{Z}\exp\left(-\frac{f(\mathbf{x})-f_{\min}}{\sigma}\right)
\end{equation}

where:
\begin{itemize}
\item $f_{\min}$  is the current minimum 
\item $\sigma$ is the standard deviation
\item $Z$ normalize the state.
\end{itemize}

The algorithm evaluates $f$ at all points in the grid $\mathbf{x} \in \mathcal{G}$ and constructs a quantum state:

\begin{equation}
    |\psi\rangle = \sum_{\mathbf{x} \in \mathcal{G}} \sqrt{d(\mathbf{x})} |\mathbf{x}\rangle
\end{equation}

where the probability amplitudes are derived from the function values:

\begin{equation}
    d(\mathbf{x}) \propto \frac{\exp\left(-\frac{f(\mathbf{x}) - f_{\text{min}}}{\sigma}\right)}{Z}
\end{equation}
The amplitude encoding follows the Boltzmann distribution that gives the probability that a system will be in a certain state as a function of that state's energy and the temperature of the system.:

\begin{equation}
    p(\mathbf{x}) \propto \exp\left(-\frac{f(\mathbf{x})}{kT_{\text{eff}}}\right)
\end{equation}

where $kT_{\text{eff}} = \sigma$ and the ground state energy is shifted by $f_{\text{min}}$.

The quantum circuit measures the state $|\psi\rangle$, yielding probabilities:

\begin{equation}
    P(\mathbf{x}) = |\langle \mathbf{x}|\psi\rangle|^2
\end{equation}

These probabilities form a discrete probability distribution over the search space, with higher probabilities corresponding to more promising regions.

We identify the most promising regions as those comprising the top 25\% probability mass:

\begin{equation}
\Omega_h^{(k)} = \{\mathbf{x} | P(\mathbf{x}) \geq P_{75}\}
\end{equation}

The new bounds for each dimension are computed by restricting to the projection of $\Omega_h^{(k)}$, where the points represent the decimal encoding of quantum states corresponding to the selected probability amplitudes. This yields the refined search domain:

\begin{equation}
[l_i^{(k+1)}, u_i^{(k+1)}] = \left[\max(l_i, \min x_i), \min(u_i, \max x_i)\right]_{\mathbf{x}\in\Omega_h^{(k)}}
\end{equation}
This contracts the search space to the hyperrectangle enclosing the high-probability region.

A classical optimization routine is then applied within the refined bounds to determine the solution.

\begin{equation}
\mathbf{x}_l^{(k+1)} = \arg\min_{\mathbf{x} \in [\mathbf{l}^{(k+1)}, \mathbf{u}^{(k+1)}]} f(\mathbf{x})
\end{equation}

The algorithm terminates when:
\begin{itemize}
\item The search space contraction becomes insignificant ($|\mathbf{u}^{(k+1)} - \mathbf{l}^{(k+1)}| < \delta$)
\item The maximum iteration count is reached ($k = K_{\max}$)
\item The quantum distribution becomes overly concentrated
\end{itemize}

The progressive contraction of the search space ensures:

\begin{equation}
\lim_{k \to \infty} \text{Vol}(\Omega^{(k)}) = 0
\end{equation}

with probability density concentrating around global minima.

\begin{exemple}
We consider a 2D function optimization with:
\begin{itemize}
\item Global bounds: $x_1 \in [-5,5]$, $x_2 \in [-10,10]$
\item Current promising region $\Omega_h^{(k)}$: $x_1 \in [-2.1,1.8]$, $x_2 \in [3.5,7.2]$
\end{itemize}

The updated bounds become:
\begin{itemize}
\item For $x_1$: $[\max(-5,-2.1), \min(5,1.8)] = [-2.1,1.8]$
\item For $x_2$: $[\max(-10,3.5), \min(10,7.2)] = [3.5,7.2]$
\end{itemize}

This strategy ensures:
\begin{itemize}
\item Progressive contraction of the search space
\item Bounds never exceed the original limits
\end{itemize}
\end{exemple}
\section{Experimental Analysis}
The proposed quantum-classical hybrid approach demonstrates compelling theoretical advantages, although its practical implementation requires careful consideration of several key factors. The simulation results are obtained using \texttt{Python 3.10, NumPy 1.24, and Qiskit 1.0.0}. However, implementing amplitude encoding on real hardware remains challenging, particularly when using a large number of qubits, as implementing numerous gates becomes impractical. Consequently, the results presented here utilize a limited number of qubits for each dimension, especially as the problem dimension increases.
Nevertheless, our method's design explicitly accounts for this constraint, offering a distinct advantage over:
\begin{itemize}
    \item Quantum approaches requiring many qubits on dense grids (e.g., VQE)
    \item Classical methods needing extensive sampling for higher accuracy
\end{itemize}
Although a low qubit count may pose challenges only for very large domains, the method remains highly promising in most scenarios.

\begin{figure}[H]
  \centering
  \begin{subfigure}[b]{0.32\textwidth}  % Larghezza ridotta per 3 figure
    \includegraphics[width=\textwidth]{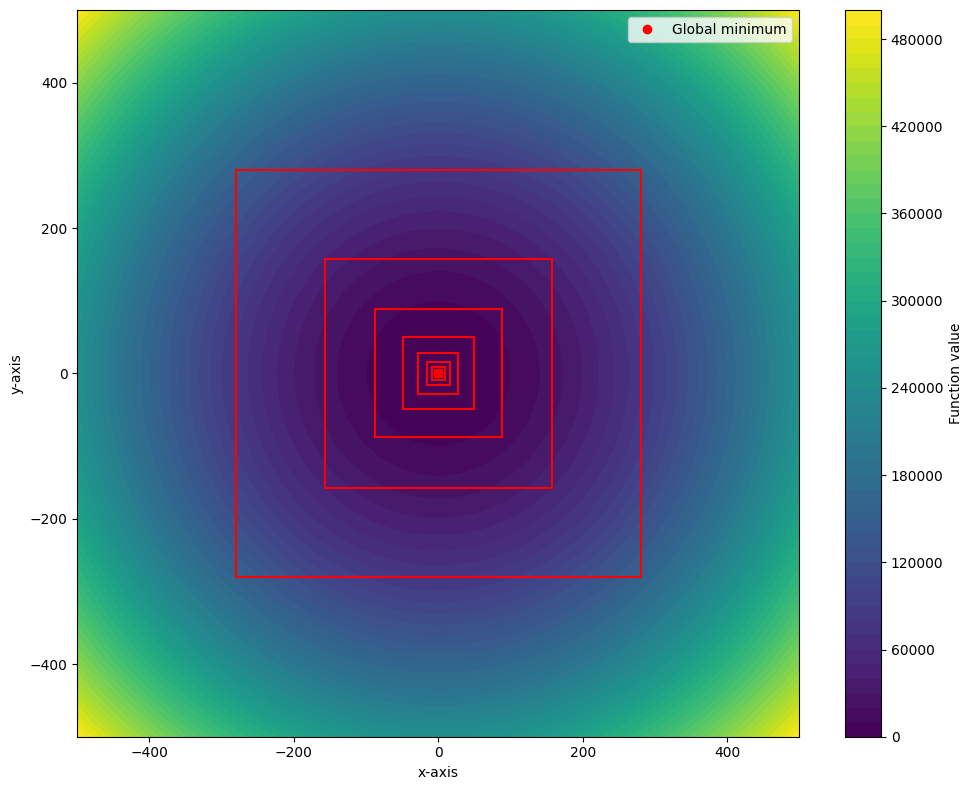}
    \caption{}
    \label{fig:sub1}
  \end{subfigure}
  \hfill
  \begin{subfigure}[b]{0.32\textwidth}
    \includegraphics[width=\textwidth]{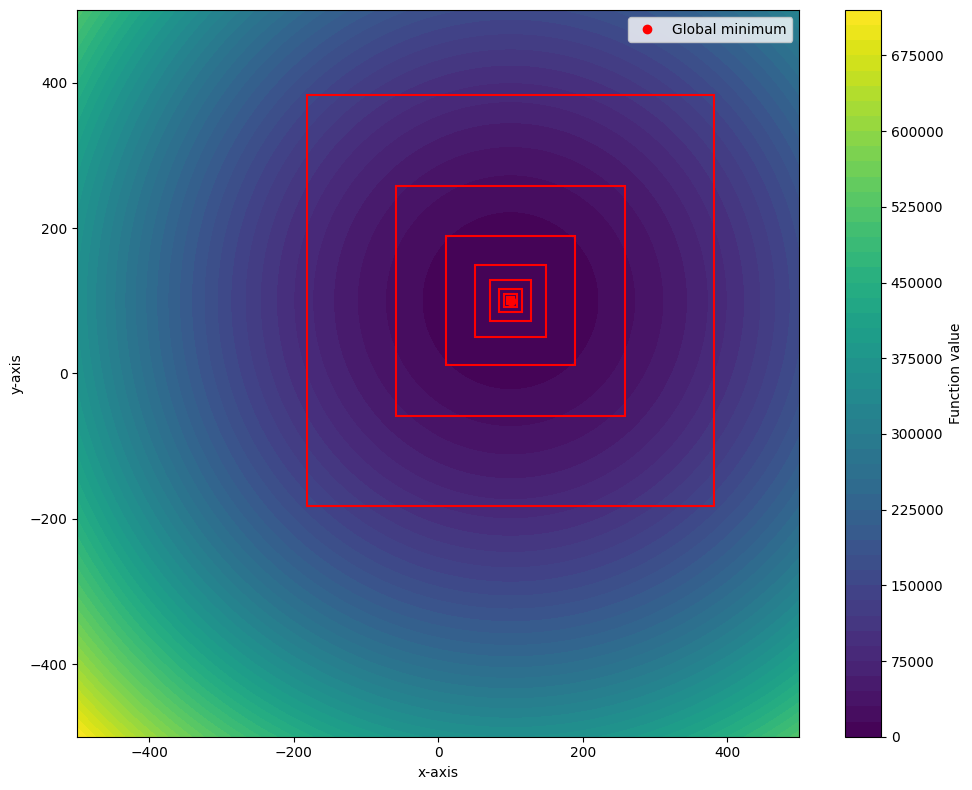}
    \caption{}
    \label{fig:sub2}
  \end{subfigure}
  \hfill
  \begin{subfigure}[b]{0.32\textwidth}
    \includegraphics[width=\textwidth]{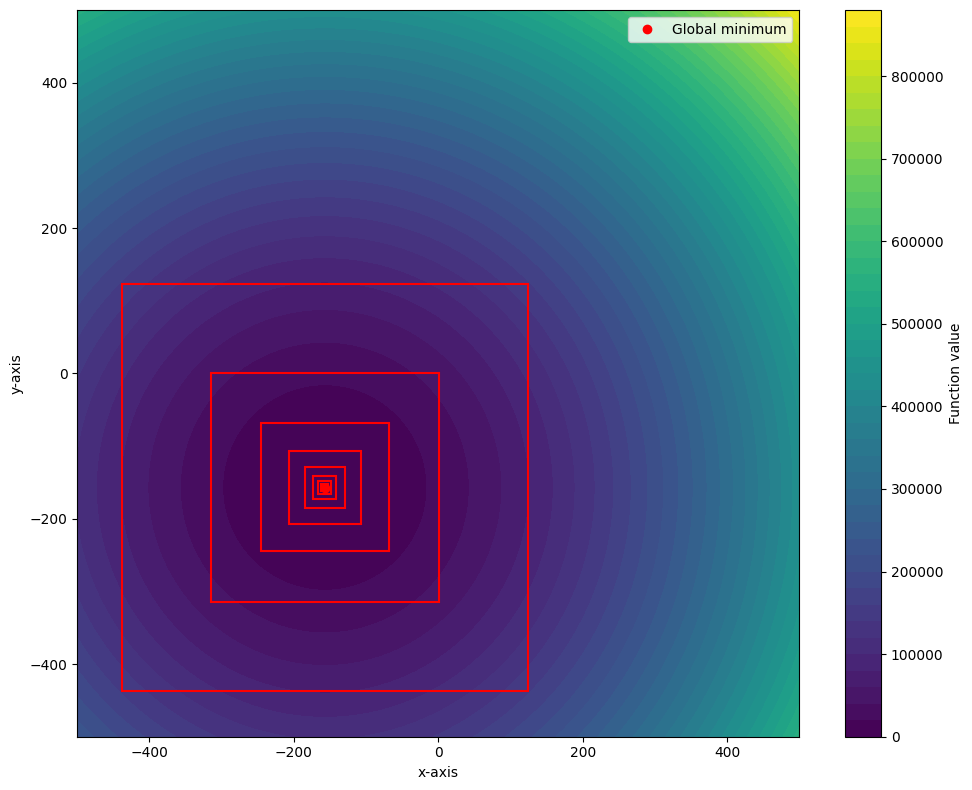}  % Sostituisci con il nome del file
    \caption{}  % Personalizza la didascalia
    \label{fig:sub3}
  \end{subfigure}
\caption{\centering
Progressive refinement of the search space on the 2D Sphere function for three different global minima: 
(a) minimum at $(0, 0)$, 
(b) minimum at $(100, 100)$, and 
(c) minimum at $(-50\pi, -50\pi)\approx(-157.08, -157.08)$. 
In each plot, the red dot marks the true global minimum $(x^*, y^*)$, while the red rectangles represent the bounding boxes at successive iterations of the quantum search algorithm.
}
  \label{fig:global}
\end{figure}

To visualize the behavior of the search process, we consider the 2D Sphere function:
\begin{equation*}
    f(x,y)=(x-x^*)^2 + (y-y^*)^2
\end{equation*}
We display three different sequences of rectangular bounding boxes that simulate the iterative contraction of the search domain. In each figure, the red rectangles represent the bounds at successive iterations, and the red dot marks the true minimum. As the iterations proceed, the bounds become increasingly narrow and concentrate around the global optimum. \\

Tests on result accuracy were performed on standard optimization benchmark functions, detailed below along with the algorithm's performance results. The classical optimizer used is \texttt{scipy.optimize.minimize} with the default \texttt{L-BFGS-B} method (a quasi-Newton bounded optimization algorithm).

\begin{figure}[H]
  \centering
  \begin{subfigure}[b]{0.32\textwidth}  % Larghezza ridotta per 3 figure
    \includegraphics[width=\textwidth]{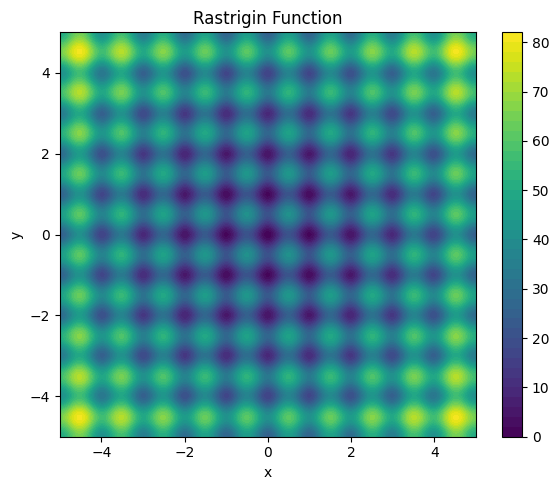}
    \caption{}
    \label{fig:sub1}
  \end{subfigure}
  \hfill
  \begin{subfigure}[b]{0.32\textwidth}
    \includegraphics[width=\textwidth]{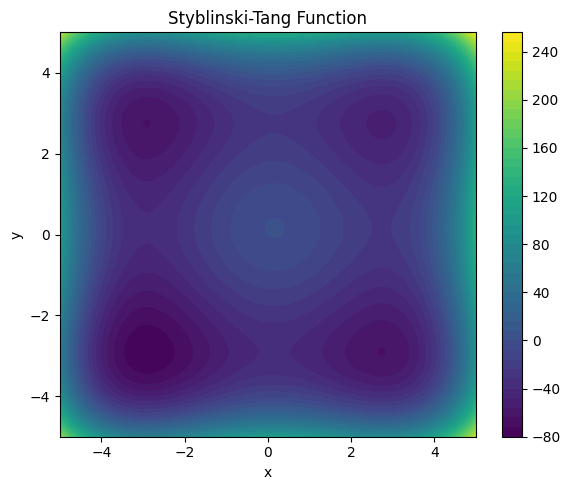}
    \caption{}
    \label{fig:sub2}
  \end{subfigure}
  \hfill
  \begin{subfigure}[b]{0.32\textwidth}
    \includegraphics[width=\textwidth]{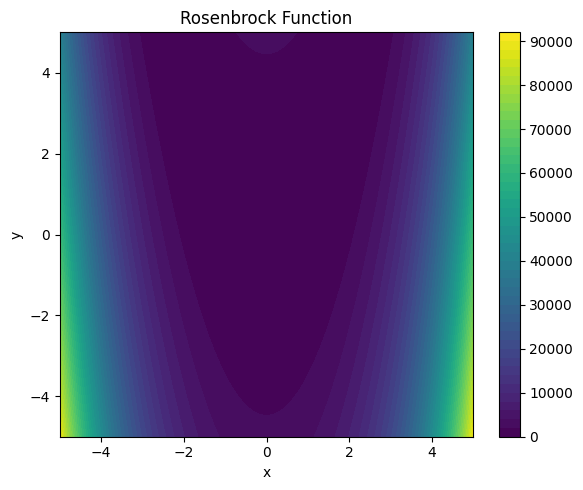}  % Sostituisci con il nome del file
    \caption{}  % Personalizza la didascalia
    \label{fig:sub3}
  \end{subfigure}
  \caption{Testing optimization algorithms on benchmark functions. Experimental results highlight performance and robustness.}
  \label{fig:global}
\end{figure}

\paragraph{Rastrigin Function}
The Rastrigin function (Figure 2(a)) is defined as follows:
\begin{equation}
    f(\mathbf{x}) = 10n + \sum_{i=1}^{n} \left[x_i^2 - 10 \cos(2\pi x_i)\right],
\end{equation}
where the domain is $[-5.12,5.12]^n$ and the global minimum is in \( \mathbf{x} = \mathbf{0} \), with \( f(\mathbf{0}) = 0 \).\\
As shown in Table 1, the algorithm produces virtually exact results when applied to the Rastrigin function in simulation.

\begin{table}[H]
\centering

\begin{tabular}{@{}rlllSS@{}}
\toprule
\textbf{Dim} & \textbf{Config} & \textbf{Found Point} & \textbf{Result} & \textbf{Real Minimum} & \textbf{Abs. Error} \\
\midrule
2  & 5 qubits & {[}0.00, 0.00{]}                                       & 0.00 & 0.00 & \num{0.00e+00} \\
3  & 4 qubits & {[}0.00, 0.00, 0.00{]}                                 & 0.00 & 0.00 & \num{0.00e+00} \\
5  & 3 qubits & {[}0.00, 0.00, 0.00, 0.00, 0.00{]}                     & 0.00 & 0.00 & \num{0.00e+00} \\
8  & 2 qubits & {[}0.00, 0.00, 0.00, \dots, 0.00{]}   & 0.00 & 0.00 & \num{0.00e+00} \\
\bottomrule
\end{tabular}
\caption{Rastrigin function results.}
\end{table}

\paragraph{Styblinski--Tang Function}
The Styblinski-Tang function (Figure 2(b)) is defined as follows:
\begin{equation}
    f(\mathbf{x}) = \frac{1}{2} \sum_{i=1}^{n} \left(x_i^4 - 16x_i^2 + 5x_i\right),
\end{equation}
where the domain is $[-5,5]^n$ and the global minimum is \( x_i \approx -2.903534 \), with \( f(\mathbf{x}) \approx -39.166n \).

\begin{table}[H]
\centering

\begin{tabular}{@{}rlllSS@{}}
\toprule
\textbf{Dim} & \textbf{Config} & \textbf{Found Point} & \textbf{Result} & \textbf{Real Minimum} & \textbf{Abs. Error} \\
\midrule
2  & 5 qubits & {[}-2.90, -2.90{]}                                                           & -78.33  & -78.33  & \num{1.31e-04} \\
3  & 4 qubits & {[}-2.90, -2.90, -2.90{]}                                                    & -117.50 & -117.50 & \num{1.97e-04} \\
5  & 3 qubits & {[}-2.90, -2.90, -2.90, -2.90, -2.90{]}                                      & -195.83 & -195.83 & \num{3.29e-04} \\
8  & 2 qubits & {[}-2.90, -2.90, -2.90, \dots, -2.90{]}                  & -313.33 & -313.33 & \num{5.26e-04} \\
\bottomrule
\end{tabular}
\caption{Styblinski–Tang function results}
\end{table}

\paragraph{Rosenbrock Function}
The Rosenbrock function (Figure 2(c)) is 
\begin{equation}
    f(\mathbf{x}) = \sum_{i=1}^{n-1} \left[100(x_{i+1} - x_i^2)^2 + (1 - x_i)^2\right],
\end{equation}
where the domain is $\mathbb{R}^n$. The global minimum in \( \mathbf{x} = (1, \dots, 1) \), with \( f(\mathbf{x}) = 0 \).

\begin{table}[H]
\centering
\begin{tabular}{@{}rlllSS@{}}
\toprule
\textbf{Dim} & \textbf{Config} & \textbf{Found Point} & \textbf{Result} & \textbf{Real Minimum} & \textbf{Abs. Error} \\
\midrule
2  & 5 qubits & {[}1.00, 1.00{]}                         & 0.00 & 0.00 & \num{2.00e-15} \\
3  & 4 qubits & {[}1.00, 1.00, 1.00{]}                   & 0.00 & 0.00 & \num{4.80e-13} \\
\bottomrule
\end{tabular}
\caption{Rosenbrock function results ($[-500,500]^n$ domain)}
\end{table}
Table 3 shows highly accurate results in an extensive domain ($[-500,500]^n$), demonstrating a significant advantage over classical methods. However, increasing the dimensionality without proportionally increasing the number of qubits may compromise both the solution accuracy and convergence, potentially leading the search toward suboptimal regions. To maintain performance without additional qubits in higher dimensions, domain restriction becomes essential, as evidenced by the results presented in Table 4.

\begin{table}[H]
\centering
\begin{tabular}{@{}rlllSS@{}}
\toprule
\textbf{Dim} & \textbf{Config} & \textbf{Found Point} & \textbf{Result} & \textbf{Real Minimum} & \textbf{Abs. Error} \\
\midrule
5  & 3 qubits & {[}1.00, 1.00, 1.00, 1.00, 1.00{]}                   & 0.00 & 0.00 & \num{1.70e-14} \\
8  & 2 qubits & {[}1.00, 1.00, 1.00, 1.00, 1.00{]}                   & 0.00 & 0.00 & \num{1.70e-14} \\
\bottomrule
\end{tabular}
\caption{Rosenbrock function results ($[-10,10]^n$ domain domain)}
\end{table}

\subsection{Classical vs. Quantum Comparison
}

This section presents a performance comparison between the QAGS algorithm and an Adaptive Grid Search. The results highlight significant advantages, particularly in memory efficiency, of the quantum-classical hybrid approach. All tests were conducted on the sphere function.

\begin{equation}
    f(\mathbf{x}) = \sum_{i=1}^{d} x_i^2 ,
\end{equation}

Table 5 shows that when both classical and quantum approaches achieve convergence, the quantum method shows significantly reduced execution times and memory usage. These results were obtained for a restricted domain of $[-5,5]^d$.

\begin{table}[H]
\centering
\label{tab:comparison_results}
\begin{tabular}{llrrrrrr}
\toprule
\multirow{2}{*}{Dim} & \multicolumn{2}{c}{Time (s)} & \multicolumn{2}{c}{Memory (MB)} & \multicolumn{2}{c}{Solution Value} \\
  & Quantum & Classic & Quantum & Classic & Quantum & Classic \\
\midrule
2 &  0.18 & 0.33 & 920.62 & 2956.47 & 0.00 & 0.00 \\
5 &  0.63 & 1.46 & 920.62 & 2956.47 & 0.00 & 0.01 \\
7 & 0.29 & 0.94 & 920.62 & 2956.51 & 0.00 & 0.00 \\
8 &  1.09 & 5.94 & 920.64 & 2956.51 & 0.00 & 0.00 \\
10 &  17.90 & 164.31 & 920.64 & 7654.20 & 0.00 & 0.00 \\
\bottomrule
\end{tabular}
\caption{Comparative Optimization Results for $[-5,5]^d$}
\end{table}

Figure 3 shows the comparative memory usage both methods. Figure 4 shows that the quantum approach demonstrates variable performance scaling with dimensionality, while classical methods exhibit more consistent behavior in lower dimensions but progressively worse scaling in higher-dimensional spaces.

\begin{figure}[H]
  \centering
    \includegraphics[width=0.8\textwidth]{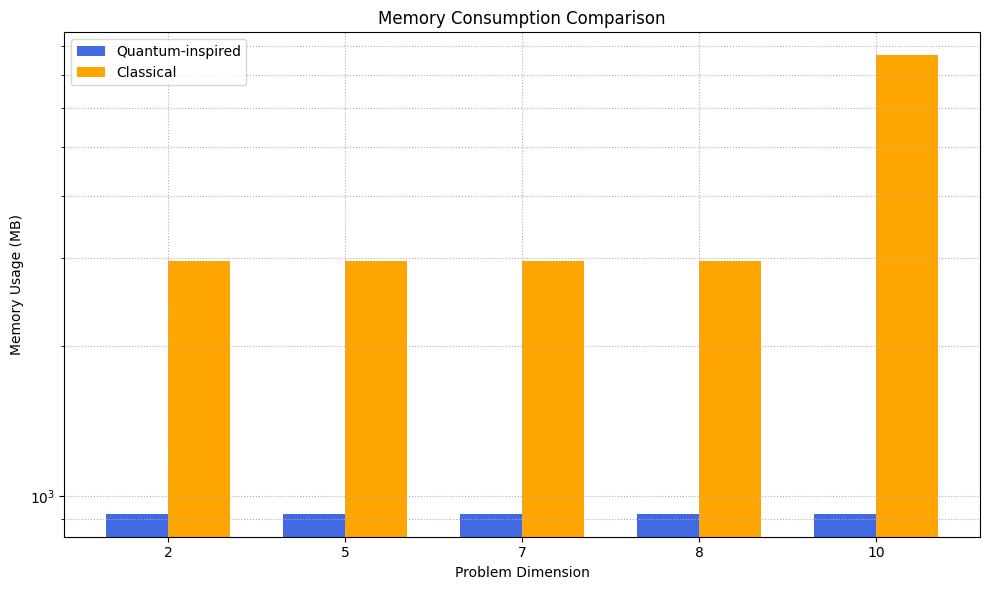}
    \caption{Memory usage distribution in hybrid-quantum computation (domain $[-5,5]^d$).}
\end{figure}

\begin{figure}[H]
  \centering
    \includegraphics[width=1\textwidth]{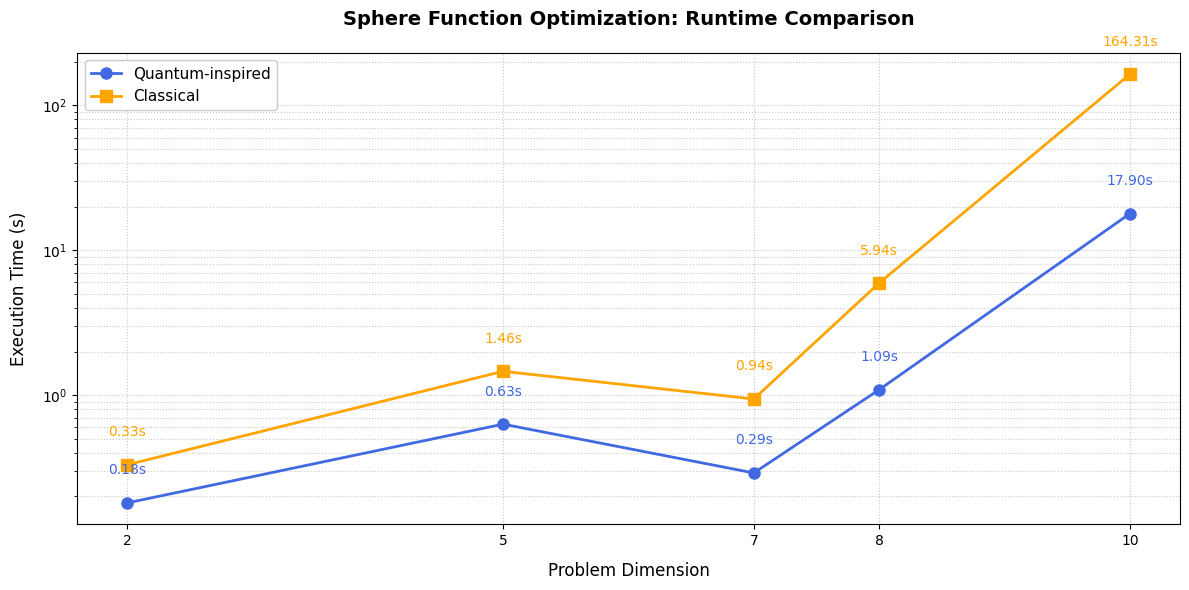}
    \caption{ Comparative performance analysis for Sphere function optimization (domain $[-5,5]^d$).}
\end{figure}

We now examine whether this performance trend persists when expanding the domain range. Table 6 presents the classical versus quantum benchmark results for the Sphere function evaluated in the extended domain $[-500,500]^d$.

\begin{table}[H]
\centering
\label{tab:comparison_results_new}
\begin{tabular}{llrrrrrr}
\toprule
\multirow{2}{*}{Dim} & \multicolumn{2}{c}{Time (s)} & \multicolumn{2}{c}{Memory (MB)} & \multicolumn{2}{c}{Solution Value} \\
  & Quantum & Classic & Quantum & Classic & Quantum & Classic \\
\midrule
2 &  0.09 & 0.30 & 402.51 & 142.59 & 1.62e-27 & 0.05 \\
5 &  0.30 & 0.11 & 407.68 & 635.05 & 0.00 & 0.00 \\
7 & 0.32 & 0.91 & 407.70 & 175.76 & 0.00 & 0.00 \\
8 &  1.04 & 4.98 & 417.31 & 307.67 & 0.00 & 0.00 \\
10 &  17.41 & 151.94 & 640.95 & 5204.81 & 0.00 & 0.00 \\
\bottomrule
\end{tabular}
\caption{Comparative Optimization Results for Sphere Function (Domain [-500,500]$^d$)}
\end{table}

Our analysis first examines the classical memory behavior, which remains approximately constant across domain variations. This stability arises from from the adaptive grid point selection strategy implemented to address computational constraints. Specifically, the classical approach was modified to reduce grid density while maintaining convergence guarantees, as excessive grid points (constrained by the curse of dimensionality) previously caused runtime failures. This optimization creates an apparent memory advantage over the quantum approach in lower dimensions.

However, the fundamental quantum advantage becomes evident at higher dimensionality. With dimension d = 10 comparative measurements show that the quantum approach requires 87.7\% less memory than the classical implementation. 

\begin{figure}[H]
  \centering
    \includegraphics[width=0.8\textwidth]{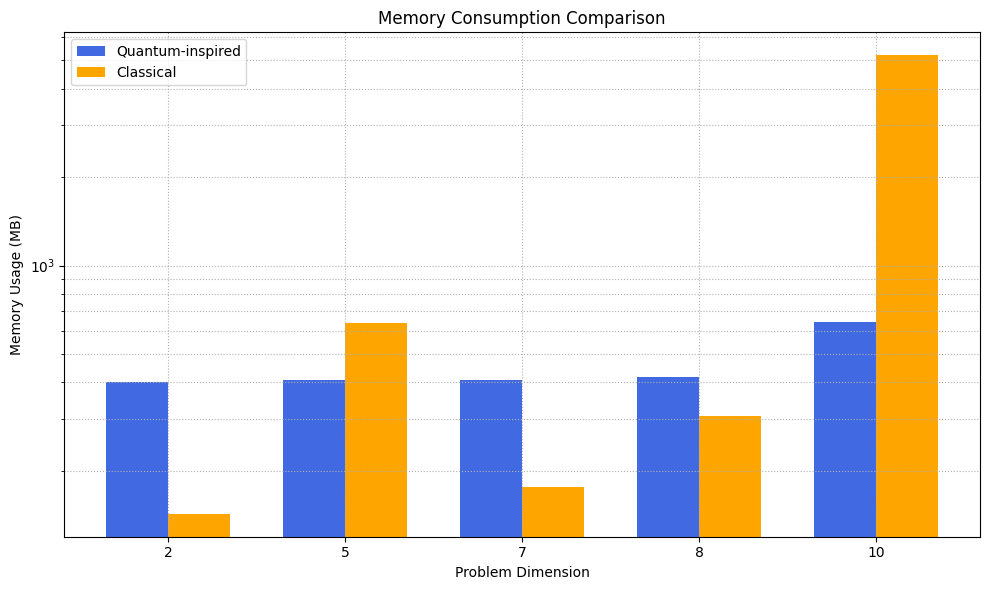}
    \caption{Memory usage distribution in hybrid-quantum computation (domain $[-500,500]^d$).}
\end{figure}

The same behavior is observed in runtime performance, as the reduction of grid points in certain dimensions may skew the results. However, the data clearly show that for dimension 10, the quantum approach achieves a 88.54\% reduction in computation time compared to the classical method.

\begin{figure}[H]
  \centering
    \includegraphics[width=1\textwidth]{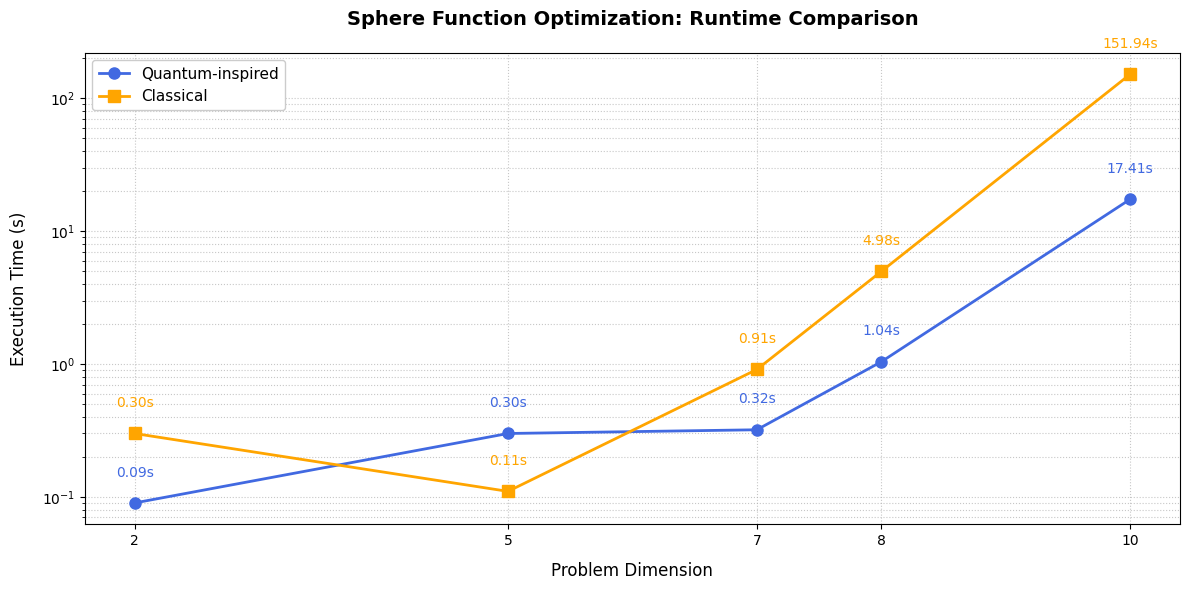}
    \caption{Comparative performance analysis for Sphere function optimization (domain $[-500,500]^d)$.}
\end{figure}

\section{Conclusion}
This study introduces a novel methodology for continuous multivariate function optimization, designed to address inherent limitations in classical optimization techniques and could have a practical quantum hardware implementations. While the presented results demonstrate significant promise, they do not incorporate potential errors and noise arising from real-world quantum gate operations. Nevertheless, our primary objective has been to establish that emerging quantum scientific computing approaches can surpass classical methods, particularly in mitigating the curse of dimensionality that plagues high-dimensional optimization problems.

The proposed approach makes substantive contributions to the field of mathematical optimization by providing a robust theoretical foundation capable of addressing complex problems while enabling new interdisciplinary applications. The methodology demonstrates particular effectiveness in scenarios where traditional methods face computational bottlenecks due to dimensional scaling. Notably, our results reveal consistent quantum advantages in both memory efficiency (up to 87.7\% reduction) and computational speed (88.54\% faster convergence) for problems beyond ten dimensions, even when accounting for domain expansion effects.

This work opens several important research directions, including the development of noise-resilient implementations and hardware-specific optimizations that could bridge the current gap between theoretical potential and practical quantum advantage. The framework establishes a crucial stepping stone toward solving optimization challenges that remain intractable for purely classical approaches.

\newpage
\nocite{*}
\bibliography{bibliography}

\end{document}